\documentclass[twocolumn,showpacs,preprintnumbers,amsmath,amssymb]{revtex4}

\usepackage{graphicx}
\usepackage{dcolumn}
\usepackage{bm}

\begin{document}


\title{Channel Formation and Intermediate Range Order in Sodium Silicate Melts and Glasses}

\author{A.\ Meyer$^1$}\email{ameyer@ph.tum.de}
\author{J.\ Horbach$^2$}\email{horbach@uni-mainz.de}
\author{W.\ Kob$^3$}
\author{F.\ Kargl$^1$} 
\author{H.\ Schober$^4$}

\affiliation{$^1$ Physik--Department E\,13, Technische Universit\"at M\"unchen, 85747 
Garching, Germany}
\affiliation{$^2$ Institut f\"ur Physik, Johannes Gutenberg--Universit\"at Mainz, 
             55099 Mainz, Germany}
\affiliation{$^3$ Laboratoire des Verres, Universit\'e Montpellier II, 34095 Montpellier, 
             France}
\affiliation{$^4$Institut Laue--Langevin, 38042 Grenoble, France}

\date{\today}

\begin{abstract}
We use inelastic neutron scattering and molecular dynamics simulation
to investigate the interplay between the structure and the fast sodium
ion diffusion in various sodium silicates.  With increasing temperature
and decreasing density the structure factors exhibit an emerging prepeak
around 0.9\,\AA$^{-1}$.  We show, that this prepeak has its origin in the
formation of sodium rich channels in the static structure.  The channels
serve as preferential ion conducting pathways in the relative immobile
Si--O matrix.  On cooling below the glass transition this intermediate
range order is frozen in.
\end{abstract}

\pacs{61.20.-p,61.20.Ja,61.12.-q}

\maketitle

Alkali silicates are a paradigm for multicomponent glass forming
systems in which even at typical melt temperatures the mobility of the
alkali ions exceeds that of the Si--O network by orders of magnitude
\cite{JoBB51,GuKi66,BrFr88}.  Although such a high mobility of the alkali
ions should be reflected in the underlying structure, up to now, it is
still unclear how the alkali atoms are built into the Si--O network.
A distribution where an increasing alkali content causes increasing
homogenous disruptions of the SiO$_4$ tetrahedal network of pure silica
\cite{Zac32,WaBi38} does not agree with the observed drastic difference
in mobility.  In addition, this distribution is in conflict with the
highly non--linear dependence of the viscosity on alkali concentration
\cite{MaSS83,KnDS94}.  These observations lead to ideas that propose
the existence of preferential ion conducting pathways in a Si--O matrix
\cite{AnCT82,Gre85}.

Here, we present evidence for the existence of sodium diffusion
channels in the static structure. Using neutron scattering experiments
and molecular dynamics simulations on various sodium silicate melts
and glasses, we investigate the structure that provides the mobility of the
alkali ions. We demonstrate the accuracy of the molecular dynamics model by
our neutron data. Then we make use of the detailed information obtained
by molecular dynamics to elucidate the observed intermediate range order
as seen by neutron scattering. 

Sodium di- (NS2), tri- (NS3) and tetra- (NS4) silicate glasses were
synthesized from ultrapure Na$_2$CO$_3$ and SiO$_2$ powders by fusion
at 1500\,K.  For the inelastic neutron scattering experiments glassy
samples were encapsulated in Pt cells giving an annular sample geometry
of 40\,mm in height, 22.5\,mm in diameter and a 1.25\,mm wall thickness.
Measurements were performed at room temperature and on liquid samples at
temperatures between 1200\,K and 1600\,K on the neutron time--of--flight
spectrometer IN\,6 of the Institut Laue-Langevin in Grenoble.  An incident
neutron wavelength of $\lambda\!=\!5.9$\,\AA$^{-1}$ yielded an energy
resolution of $\delta E=50\,\mu\mbox{eV}$ (FWHM) and an accessible
momentum transfer at zero energy transfer of $q=0.2-1.75$\,\AA$^{-1}$.
The raw data reduction consisted of a normalization to a vanadium standard
and a correction for self absorption and container scattering.

At 1600\,K the structural relaxation of the Si--O network
for the investigated sodium silicates is on a nanosecond scale
\cite{MeSD02,KnDS94} -- too slow to be resolved on IN\,6.  Therefore,
the dynamic structure factor $S(q,\omega)$, depending on momentum $q$
and energy transfer $\hbar\omega$, exhibits a strong elastic line with an
elastic structure factor $S(q,\omega\!=\!0)$ that is in good approximation
the static structure factor times the Debye--Waller factor $f(q)$.
Whereas scattering on the Si and O atoms is exclusively coherent,
scattering on sodium is coherent and incoherent.  The incoherent
scattering reflects itself in a flat background in $S(q,\omega\!=\!0)$.

The potential that was used for the molecular dynamics simulations
is a slight modification of the pair potential by Kramer {\it et
al.}~\cite{KrMS91} which is based on {\it ab initio} calculations.
More details on this potential can be found in Ref.~\cite{HoKB99}.
The simulations were done at various densities for systems of $N=8064$
particles ($N_{\rm Si}=2016$, $N_{\rm Na}=1344$, and $N_{\rm O}=4704$).
At the densities $2.2$\,g/cm$^3$ and $2.37$\,g/cm$^3$ systems at the
temperature $T=2100$\,K, i.e.\ in the liquid state, were equilibrated
for 3.3\,ns followed by microcanonical production runs of the same
length (using the velocity form of the Verlet algorithm with a time
step of 1.6\,fs).  In each case two independent runs were done in
order to improve the statistics.  The equilibrated systems at the
density $\rho=2.37$\,g/cm$^3$ were also quenched to a glass state
at $T=300$\,K using a cooling rate of about $10^{12}$\,K/s.  In the
latter simulation the pressure was kept constant at ambient pressure
which yielded a density of $2.46$\,g/cm$^3$ at 300\,K.  Note that
the densities $\rho=2.2$\,g/cm$^3$ and $\rho=2.46$\,g/cm$^3$ are
close to the experimental densities at $T=2100$\,K and $T=300$\,K,
respectively~\cite{MaSS83}.

The central quantity for the determination of the elastic structure
factor from the simulation is the time--dependent dynamic structure
factor defined by
\begin{equation}
  S(q,t) = \frac{N}{\sum_{\alpha} N_{\alpha} b_{\alpha}^2}
  \sum_{\alpha, \beta} b_{\alpha} b_{\beta} S_{\alpha \beta}(q,t)
  \quad \alpha, \beta \in \{ {\rm Si, Na, O }\}
 \label{eq1}
\end{equation}
with
\begin{equation}
  S_{\alpha \beta}(q,t)  =  \frac{1}{N}
  \sum_{k=1}^{N_{\alpha}}
  \sum_{l=1}^{N_{\beta}} \langle 
  \exp [i {\bf q} \cdot ({\bf r}_k(t) - {\bf r}_l(0)) ] \rangle \quad ,
\end{equation}
${\bf r}_i (t)$ being the position vector of particle $i$ at time $t$.
In Eq.~(\ref{eq1}) $b_{\alpha}$ denotes the experimental coherent
scattering lengths of particle species $\alpha$.  The elastic structure
factors were estimated by the equation $S(q,\omega=0)=f(q) S(q)$
(with $S(q)\equiv S(q,t=0)$) where the Debye--Waller factor $f(q)$
was determined by fitting in the  $\beta-$relaxation
regime the normalized quantity $F(q,t) = S(q,t)/S(q)$
to a von Schweid\-ler law (for details see Ref.~\cite{HoK02}).
Note that we have not taken into account the incoherent scattering on
sodium in the calculation of $S(q,\omega=0)$ because this contribution
is not expected to change the shape of the function $S(q,\omega=0)$
significantly.

\begin{figure}[b]
\includegraphics[width=0.9\columnwidth]{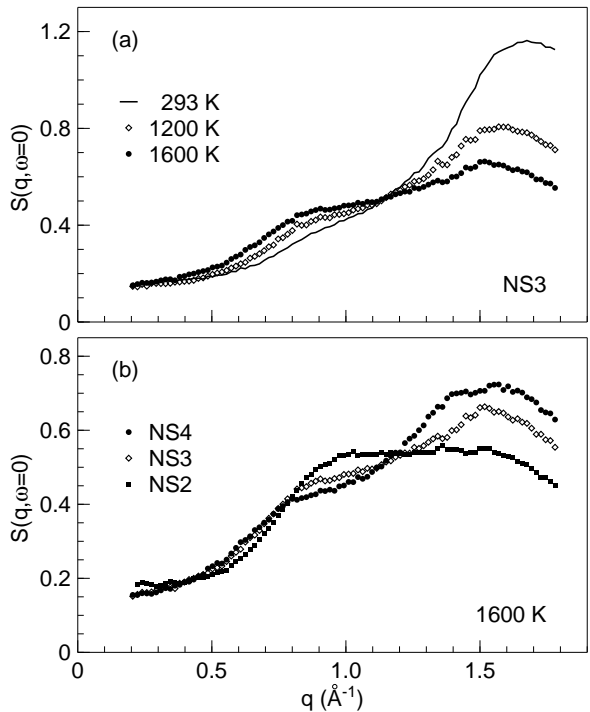}
\caption{\label{nsqns} 
Elastic structure factor as seen by inelastic neutron scattering:
Note the emerging prepeak at $\simeq 0.9$\,\AA$^{-1}$ with increasing temperature 
in sodium trisilicate (a) with a position that is
fairly independent of sodium concentration (b).}
\end{figure}

Fig.~\ref{nsqns}a displays the elastic structure factor of
glassy and viscous sodium trisilicate as seen by inelastic neutron
scattering.  The maximum at $\simeq\!1.7\,$\AA$^{-1}$ corresponds to
the disrupted tetrahedal Si--O network structure of alkali silicate
\cite{WaSu77,StMD95}.  With increasing temperature the elastic scattered
intensity at this wave-vector is decreasing at the expense of an
increasing inelastic scattering.  In contrast $S(q,\omega\!=\!0)$
displays an additional prepeak at $\simeq\!0.9\,$\AA$^{-1}$
which grows with increasing temperature, indicating an enhanced probability for a
correlation of atomic arrangements at distances around 6-8\,\AA.

In sodium disilicate \cite{MeSD02} and sodium tetrasilicate neutron
scattering reveals a similar behaviour: The elastic structure factors
exhibit a growing prepeak at $\simeq 0.9$\,\AA$^{-1}$ on temperature
increase.  With increasing sodium content the height of the prepeak
is increasing, whereas its position remains fairly unaffected by this
large change in composition (Fig.~\ref{nsqns}b).  We note, that for
a monoatomic system a change in the structure that is due to thermal
expansion would lead -- beside a slight shift of the peak positions
towards a smaller $q$ value -- to a decreasing intensity of the peak
height with increasing temperature.  Thus, the emerging prepeak in the
elastic structure factors as shown in Fig.~\ref{nsqns} does not seem to
be merely caused by thermal expansion but might be related to a change
in the underlying structure.  In the following, we clarify this behaviour
of the elastic structure factors by means of our molecular dynamics simulations.

\begin{figure}[b]
\includegraphics[width=0.9\columnwidth]{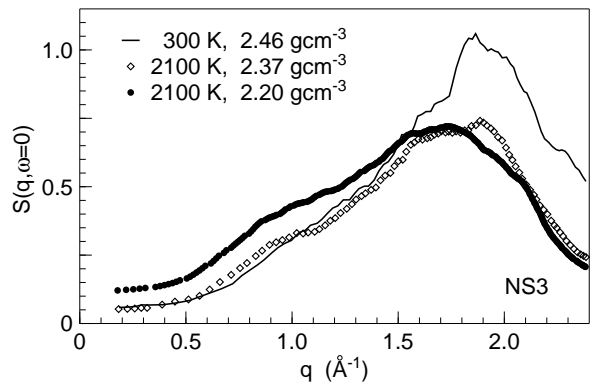}
\caption{\label{ns3mdsq} 
Elastic structure factor of sodium trisilicate as obtained by molecular
dynamics simulation (see text) weighted with the neutron scattering lengths:
Note the emerging prepeak at $\simeq 0.9$\,\AA$^{-1}$ with decreasing density.
}
\end{figure}

\begin{figure}[t]
\includegraphics[width=0.9\columnwidth]{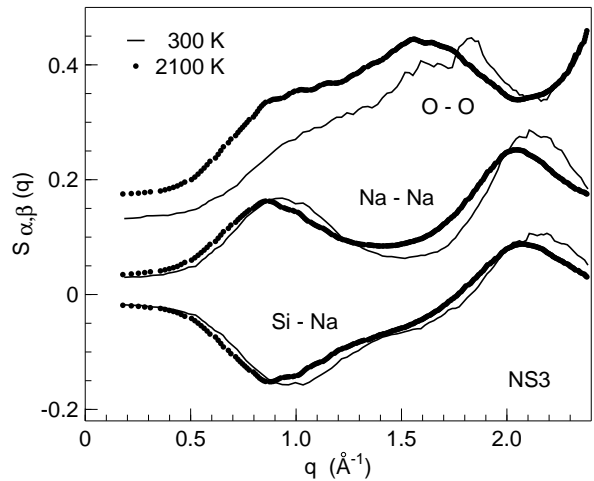}
\caption{\label{ns3part} 
Partial structure factors $S_{\rm O,O}(q)$, $S_{\rm Na,Na}(q)$, $S_{\rm Si,Na}(q)$
in glassy and liquid sodium trisilicate at experimental 
densities. $S_{\rm O,O}(q) + 0.1$ for clarity.
}
\end{figure}

Fig.~\ref{ns3mdsq} displays the ``simulated'' neutron scattering elastic
structure factor of sodium trisilicate.  At the experimental densities,
$S(q,\omega=0)$ exhibits the same behavior as in the neutron scattering
experiment~\cite{incsca}: Whereas around 1.7\,\AA$^{-1}$ the elastic
scattered intensity decreases with increasing temperature (as expected), around
0.9\,\AA$^{-1}$ a shoulder is present at $T=2100$\,K which is nearly
absent in the glass at $T=300$\,K.  Also shown in Fig.~\ref{ns3mdsq}
is the elastic structure factor at the density of $2.37$~g/cm$^3$ at
$T=2100$\,K. Here the shoulder around 0.9\,\AA$^{-1}$ has a smaller
amplitude than in the corresponding case at the experimental density.
This gives evidence that the possible structural changes leading to
a more pronounced appearance of the feature around 0.9\,\AA$^{-1}$ in
$S(q,\omega=0)$ are not due to a change in temperature but due to the
change in the density. Thus, the question arises whether the shoulder
around 0.9\,\AA$^{-1}$ is related to structural features that tend
to disappear with increasing density.

The partial static structure factors $S_{\alpha \beta}(q)$ at the
experimental densities for different correlations are shown in
Fig.~\ref{ns3part}: A well--pronounced peak is present in $S_{\rm
Si,Na}(q)$ and $S_{\rm Na,Na}(q)$ at $T=2100$\,K {\it and}~at $T=300$\,K.
Thus, the structure that leads to the peak at $q_1=0.9$\,\AA$^{-1}$ does
not at all disappear with increasing density. That the feature at $q_1$
seems to be absent in $S(q,\omega=0)$ at 300\,K, is due to the fact that
$S(q,\omega=0)$ is a linear combination of six different partial structure
factors: On the one hand positive and negative contributions cancel
each other (note the negative amplitude of $S_{\rm Si,Na}(q)$ at $q_1$),
and on the other hand $S_{\rm O,O}(q)$ exhibits only a shoulder at $q_1$
which is less pronounced at $T=300$\,K.  Because oxygen is the majority
component (about 60\% of the particles in NS3) and the coherent scattering
length of oxygen is significantly larger than that of sodium and silicon
($b_{\rm O}/b_{\rm Si}\approx 1.4$ and $b_{\rm O}/b_{\rm Na}\approx
1.6$), $S_{\rm O,O}(q)$ gives the major contribution to $S(q,\omega=0)$
and thus, the changes in $S_{\rm O,O}(q)$ are the main cause for the
emerging prepeak in the experimental and simulated $S(q,\omega=0)$
with increasing temperature and decreasing density, respectively.

\begin{figure}[b]
\includegraphics[width=7.4cm]{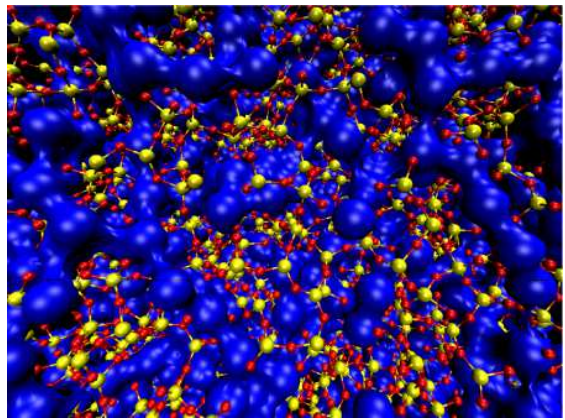}
\caption{\label{snap} 
Molecular dynamics snapshot of the structure of sodium trisilicate at 
2100\,K at the density
$\rho=2.2$\,g/cm$^3$: The blue spheres that are connected to each other
represent the Na atoms. The Si--O network is drawn by yellow (Si) and red (O) spheres
that are connected to each other by covalent bonds shown as sticks between Si
and O spheres.}
\end{figure}

The feature at $q_1$ corresponds to a distribution of sodium ions that
is not homogenous on a length scale of 6--8\,\AA. This is illustrated
by the snapshot, Fig.~\ref{snap}, which shows a molecular dynamics
configuration of NS3 at $T=2100$\,K at the density 2.2\,g/cm$^3$.  Here,
the sodium ions are represented by blue spheres that are connected to
each other by surfaces.  For clarity, the Si and O atoms are shown as
small yellow and red spheres, respectively, that are connected to each
other by covalent bonds shown as sticks. Note that the size of the latter
spheres does not mirror the real size of Si and O. The snapshot gives
a clear picture of a network of sodium channels that percolate through
the Si--O structure. The peak at $q_1$ in the static structure factor
marks the characteristic length scale of the latter network of channels.

Recently, Jund {\it et al.}~\cite{JuKJ01} have shown in a molecular
dynamics simulation of NS4 that the sodium trajectories form a network
of pockets and channels where the distance between the pockets is of
the order of 5--8~\AA. Then,  Horbach {\it et al.}~\cite{HoKB02} have
given evidence in the case of NS2 that the latter network is reflected
by the prepeak at $q_1$ in the static structure factor. Fig.~\ref{snap}
shows that the ``structure of the sodium trajectories'' is not only
seen in the dynamics as claimed by Jund {\it et al.}~\cite{JuKJ01}.
The sodium trajectories are given by paths in a quasi--static structure
of sodium diffusion channels, the lifetime of which is given by the
characteristic relaxation time of the Si--O network.  

A change of the Na$_2$O content (from NS4 to NS2) or of the density does
not significantly affect the position of the prepeak at $q_1$ as seen by
molecular dynamics simulations {\it and} neutron scattering.  The effect
that the prepeak at $q_1$ becomes more pronounced in $S(q,\omega=0)$ with
increasing Na$_2$O concentration (Fig.~\ref{nsqns}b) can be also explained
by the simulation~\cite{HoKB99}: The amplitude of the peak around
$1.7$~\AA$^{-1}$ decreases with an increasing sodium content indicating
a stronger disruption of the Si--O network if more sodium ions are
added to the system. Moreover, Na--Na correlations contribute with a
larger weight in the total structure factor and thus also the amplitude
of the prepeak at $q_1$ increases.  These findings from the simulation
explain the behavior of $S(q,\omega=0)$ as seen in Fig.~\ref{nsqns}b.

However, a change in the composition or the density is accompanied by
a significant change in the disrupted Si--O network structure. This is
indicated by the behaviour of the Si--O--Si angle: In the NS3 simulation
at the experimental densities this angle increases from 134.5$^\circ$
at $T=300$~K (i.e.~at the density of 2.46\,g/cm$^3$) to 143$^\circ$
(i.e.~at the density of 2.2\,g/cm$^3$).  Thus, the sodium diffusion
channels are sustained over a broad range in temperature and sodium
concentration because the packing of the disrupted Si--O tetrahedra is
rather ``flexible'' with respect to the Si--O--Si angle.  Of course,
there should be a lower bound of sodium concentrations below which the latter
rearrangements in the Si--O network are no longer possible.  Indeed, at
very low sodium oxide content (below about 5\,mol\%), a miscibility
gap of Na$_2$O and SiO$_2$ even at typical melt temperatures has been
reported~\cite{mazurin}.

Very recently, Lammert {\it et al.}~\cite{LKH03} have investigated
a lithium silicate melt by molecular dynamics simulation.  They have
shown that the diffusion mechanism for lithium is a kind of vacancy
diffusion. Movies that we have done from the NS3 configurations such as
Fig.~\ref{snap} reveal a similar sodium diffusion mechanism.  Thus, our
picture of diffusion channels is consistent with the vacancy diffusion
on a local length scale as proposed by Lammert {\it et al.} provided
that also in lithium silicates diffusion channels exist.
Indeed, our recent results of inelastic neutron scattering measurements
\cite{KaMe03} and molecular dynamics simulations \cite{KnHB03} on various
lithium and potassium bearing silicates indicate that preferential ion
conduction pathways are a more general feature for alkali silicates.
A neutron scattering experiment on a calcium silicate glass using Ca
isotope substitution revealed a prepeak in the Ca--Ca partial structure
factor \cite{GaEB91} at $\simeq 1.3$\,\AA$^{-1}$.  In contrast to
our findings in alkali silicates, the signal has been interpreted in
terms of a glass structure based on ordered, densily packed domains.
It remains to be seen, to what extent the existence of preferential
ion conducting pathways in the static structure shown here for sodium
silicates applies to other glass forming, oxide ion conductors.

In conclusion, inelastic neutron scattering experiments and molecular
dynamics simulations on sodium silicate melts and glasses revealed
the existence of sodium diffusion channels in the static structure.
We find, that the formation of preferential ion conducting pathways is
fairly unaffected by the sodium oxide concentration (in the studied
composition range from sodium tetrasilicate to sodium disilicate),
by a temperature change corresponding to the glass at 300\,K up to the
equilibrium melt, or a change of the density in the melt of about 8\,\%.
In sodium silicates the values of the sodium ion diffusion at typical
melt temperatures \cite{JoBB51,GuKi66,BrFr88} and the values of the
viscosities \cite{MaSS83,KnDS94} depend relatively weakly on the sodium
oxide concentration.  The formation of sodium rich channels gives an
explanation on a microscopic level for the observed macroscopic properties
of mass transport in sodium silicates.

We thank Kurt Binder, Donald B.\ Dingwell, and Winfried Petry
for their support and fruitful discussions. We are grateful to Hans Knoth
for preparing Fig.~\ref{snap}. One of us (J.~H.) was supported by the
Deutsche Forschungsgemeinschaft (DFG) grant N$^{\rm o}$ HO2231/2--1.

\end{document}